\documentclass[sigconf,screen]{acmart}

\usepackage{algorithm}
\usepackage[noend]{algpseudocode}
\usepackage{textcomp}
\usepackage[labelformat=simple]{subcaption}
\usepackage{listings}
\usepackage{xspace}
\usepackage[inline,shortlabels]{enumitem}
\usepackage{calc}
\usepackage{pifont}
\usepackage{algorithm}
\usepackage{amsfonts}
\usepackage{amsthm}
\usepackage{xspace}
\usepackage{pifont}
\usepackage{multirow}
\usepackage{enumitem}
\usepackage{color,colortbl}
\usepackage{listings}

\usepackage{subcaption}

\newcommand{\nonterm}[1]{\ensuremath{\mathit{#1}}}

\newcommand{\T}[1]{\mbox{\lstinline[basicstyle=\ttfamily,columns=fixed]^#1^}}

\usepackage{framed}
\usepackage{xcolor}

\definecolor{tcolorboxbg}{RGB}{240, 249, 255}
\definecolor{tcolorboxborder}{RGB}{34,180,255}

\newenvironment{tcolorbox}{%
  \parindent0pt\parskip\baselineskip
  \vspace{0.1cm}
  \MakeFramed{\advance\hsize-\width\FrameRestore}%
}{%
  \endMakeFramed%
  \vspace{0.1cm}
}

\newcommand{\eg}{\textit{e.g.\@}\xspace}

\newcommand{\ie}{\textit{i.e.\@}\xspace}

\def\BibTeX{{\rm B\kern-.05em{\sc i\kern-.025em b}\kern-.08em
    T\kern-.1667em\lower.7ex\hbox{E}\kern-.125emX}}

\definecolor{darkgreen}{rgb}{0,0.42,0.24}
\lstdefinelanguage{diff}{
	morecomment=[f][\color{blue}]{@@},     
	morecomment=[f][\color{red}]-,         
	morecomment=[f][\color{darkgreen}]+,       
	morecomment=[f][\color{magenta}]{---}, 
	morecomment=[f][\color{magenta}]{+++},
}

\newcommand\lt[1]{{\lstinline+#1+}} 
\renewcommand\t[1]{{\lstinline+#1+}} 
\definecolor{dkgreen}{rgb}{0,0.5,0}
\definecolor{dkred}{rgb}{0.5,0,0}
\definecolor{gray}{rgb}{0.5,0.5,0.5}
\let\origthelstnumber\thelstnumber
\makeatletter
\newcommand*\Suppressnumber{%
  \lst@AddToHook{OnNewLine}{%
    \let\thelstnumber\relax%
     \advance\c@lstnumber-\@ne\relax%
    }%
}

\newcommand*\Reactivatenumber{%
  \lst@AddToHook{OnNewLine}{%
   \let\thelstnumber\origthelstnumber%
   \advance\c@lstnumber\@ne\relax}%
}

\definecolor{LightGray}{gray}{0.9}
\definecolor{Gray}{gray}{0.8}

\hypersetup{%
 pdftitle = {our title},
 pdfkeywords = {},
 pdfauthor = {anonymous},
 bookmarksnumbered,
 bookmarksopen=false,
 urlcolor=black,
 linkcolor=black,
 citecolor=black,
 colorlinks=true,
 pdfstartview={FitH},
}

\definecolor{skcolor}{rgb}{0.1,0.7,0.8}
\definecolor{efcolor}{RGB}{255, 0, 255}
\definecolor{ybcolor}{RGB}{0,100,50}
\definecolor{trcolor}{rgb}{0.7,0.3,0.7}
\definecolor{slcolor}{rgb}{0.1,0.4,0.8}
\definecolor{aacolor}{RGB}{255, 165, 0}

\newcommand{\tool}{\textsc{Cobblestone}\xspace}

\newcommand{\proverbot}{Proverbot9001\xspace}
\newcommand{\tactician}{Tactician\xspace}

\newcommand{\coqhammer}{CoqHammer\xspace}
\newcommand{\palm}{PALM\xspace}
\newcommand{\rango}{Rango\xspace}
\newcommand{\regen}{LinearRegen\xspace}

\newcommand{\coqgym}{CoqGym\xspace}
\newcommand{\wigderson}{coq-wigderson\xspace}
\newcommand{\pnv}{PnVRocqLib\xspace}
\newcommand{\bb}{coq-bb5\xspace}
\newcommand{\coqgymtest}{CoqGym100\xspace}
\newcommand{\wigdersontest}{Wigderson100\xspace}
\newcommand{\pnvtest}{PnVRocqLib100\xspace}
\newcommand{\bbtest}{BB5100\xspace}
\newcommand{\gptfour}{GPT-4\xspace}
\newcommand{\gptthree}{GPT-3.5-turbo\xspace}
\newcommand{\claude}{Claude 3 Opus\xspace}

\newcommand{\circledletter}[2][blue]{\raisebox{-2pt}{\includegraphics[width=10pt]{./img/#2-#1.pdf}}}

\newcommand{\mypara}[1]{\smallskip \noindent \textbf{#1}}

\definecolor{codegreen}{rgb}{0,0.6,0}
\lstdefinestyle{CoqStyle}{
    backgroundcolor=\color{white},   
    commentstyle=\color{green},
    keywordstyle=\color{blue},
    numberstyle=\tiny\color{black},
    stringstyle=\color{black},
    basicstyle=\ttfamily\scriptsize,
    morekeywords={Require,Import,Proof,Qed,Lemma,Theorem,Admitted},
    breakatwhitespace=false,         
    breaklines=true,                 
    captionpos=b,                    
    keepspaces=true,                 
    numbers=left,                    
    xleftmargin=5.0ex,
    numbersep=5pt,                  
    showspaces=false,                
    showstringspaces=false,
    showtabs=false,                  
    tabsize=2
}

\acmBadgeR[https://www.acm.org/publications/policies/artifact-review-and-badging-current]{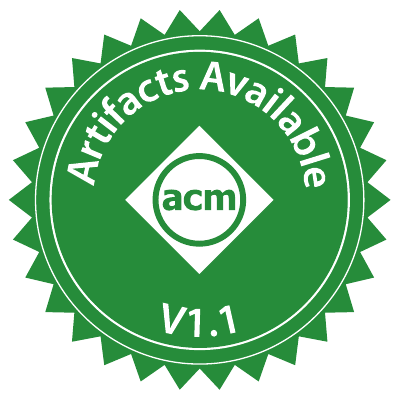}
\acmBadgeR[https://www.acm.org/publications/policies/artifact-review-and-badging-current]{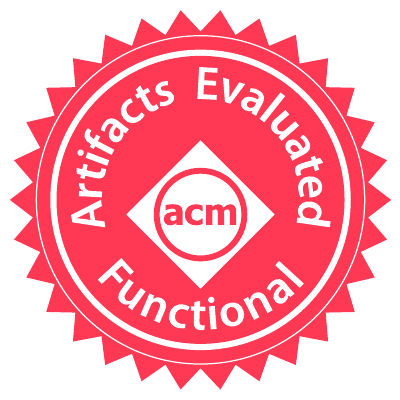}
\acmBadgeR[https://www.acm.org/publications/policies/artifact-review-and-badging-current]{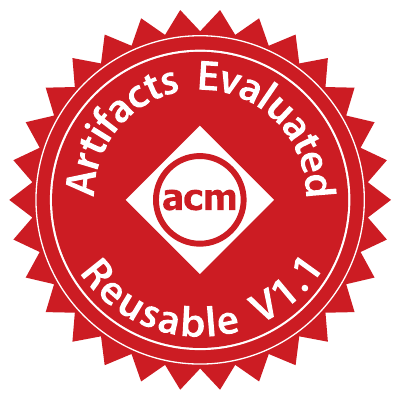}

\begin{document}

\title[Cobblestone: A Divide-and-Conquer Approach for Automating Formal Verification]{Cobblestone: A Divide-and-Conquer Approach for Automating\\Formal Verification}

\author{Saketh Ram Kasibatla}
\orcid{0009-0003-1972-7146}
\affiliation{%
  \institution{UC San Diego}
  \city{San Diego}
  \state{California}
  \country{USA}
}
\email{skasibatla@ucsd.edu}

\author{Arpan Agarwal}
\orcid{0009-0001-5193-0959}
\affiliation{%
  \institution{University of Illinois Urbana-Champaign}
  \city{Urbana-Champaign}
  \state{Illinois}
  \country{USA}
}
\email{arpan.agrawal94@gmail.com}

\author{Yuriy Brun}
\orcid{0000-0003-3027-7986}
\affiliation{%
  \institution{University of Massachusetts}
  \city{Amherst}
  \state{Massachusetts}
  \country{USA}
}
\email{brun@cs.umass.edu}

\author{Sorin Lerner}
\orcid{0000-0003-3957-0628}
\affiliation{%
  \institution{Cornell University}
  \city{Ithaca}
  \state{New York}
  \country{USA}
}
\email{sorin.lerner@cornell.edu}

\author{Talia Ringer}
\orcid{0000-0003-1854-3321}
\affiliation{%
  \institution{University of Illinois Urbana-Champaign}
  \city{Urbana-Champaign}
  \state{Illinois}
  \country{USA}
}
\email{tringer@illinois.edu}

\author{Emily First}
\orcid{0000-0002-2896-2928}
\affiliation{%
  \institution{Rutgers University}
  \city{New Brunswick}
  \state{New Jersey}
  \country{USA}
}
\email{emily.first@rutgers.edu}



\begin{abstract}

Formal verification using proof assistants, such as Coq, is an effective way of improving software quality, but 
requires significant effort and expertise.
Machine learning can automatically synthesize proofs, but such tools are able to prove only a fraction of desired software properties.
We introduce \tool, a divide-and-conquer approach for proof synthesis.
\tool uses a large language model (LLM) to generate potential proofs, uses those proofs to break the problem into simpler parts, 
automatically identifies which of those parts were successfully proven, and iterates on the remaining parts to build a correct proof that is guaranteed to be sound, despite the reliance on unsound LLMs. 
We evaluate \tool on four benchmarks of open-source Coq projects, controlling for training data leakage.
Fully automatically, \tool outperforms state-of-the-art non-LLM tools, and proves many theorems that other LLM-based tools cannot, and on many benchmarks, outperforms them.
Each \tool run costs only \$1.25 and takes 14.7 minutes, on average.
\tool can also be used with external input, from a user or another tool, providing a proof structure or relevant lemmas. 
Evaluated with such an oracle, \tool proves up to 58\% of theorems. 
Overall, our research shows that tools can make use of partial progress and external input to more effectively automate formal verification. 
\end{abstract}

\copyrightyear{2026}
\acmYear{2026}
\makeatletter
\def\@ACM@copyright@check@cc{}
\makeatother
\setcopyright{cc}
\setcctype{by}
\acmConference[ICSE '26]{2026 IEEE/ACM 48th International Conference on Software Engineering}{April 12--18, 2026}{Rio de Janeiro, Brazil}
\acmBooktitle{2026 IEEE/ACM 48th International Conference on Software Engineering (ICSE '26), April 12--18, 2026, Rio de Janeiro, Brazil}
\acmPrice{}
\acmDOI{10.1145/3744916.3773178}
\acmISBN{979-8-4007-2025-3/2026/04}



\maketitle

\pagestyle{fancy}

\section{Introduction}
\label{sec:Introduction} 

Bugs in software systems can be costly and dangerous. 
In 2022, poor software quality cost the US economy \$2.41 trillion~\cite{Krasner22}, and bugs can bring down critical, global systems~\cite{Menn24}. 
Formal verification using proof assistants, such as Coq~\cite{coq}
or Lean~\cite{Moura21}, is a promising method of improving software quality. 
Formal verification can mathematically prove the absence of entire classes of bugs, providing strong guarantees for the correctness of critical software systems.
And formal verification is highly effective: 
A study~\cite{Yang11a} of C compilers found bugs in every tested compiler, including LLVM~\cite{Lattner04} and GCC~\cite{Stallman12}, but not in the formally verified (in Coq) portions of CompCert~\cite{Leroy09}.

But, formal verification requires specifying desired properties, writing mathematical proofs of the properties, and machine checking those proofs using the proof assistant.
Writing these proofs requires significant expertise and manual effort.
For example, the proofs verifying CompCert are 8~times longer than the functional code~\cite{Leroy06}, and even small changes to the software can require heavy proof editing~\cite{Ringer21}.
While hundreds of large software systems have been verified~\cite{Ringer21}, like the sel4 microkernel~\cite{Klein09, Murray13} and the CakeML compiler~\cite{kumar2014cakeml}, and formal verification has seen industrial success, e.g., at Airbus France~\cite{Souyris14}, Google, and Mozilla~\cite{Erbsen19, Jacobs20}, most software today is not verified due to the high manual cost. 

Recent work has aimed to reduce the cost of formal verification by using machine learning to synthesize proofs~\cite{Sanchez20, First20oopsla, First22icse, Sanchez-Stern23toplas, Paliwal20, Yang19, First23fse}.
However, these approaches, even when combined with an SMT-solver-based hammer~\cite{Czajka18} can only prove one third of the desired properties on a large benchmark of open-source Coq projects~\cite{First22icse}. The advent of large language models (LLMs) has improved performance in this space~\cite{Lu24, thompson2025rango, thakur2024an}. 

In this paper, we aim to occupy the happy medium between two common approaches to proof synthesis. 
Some work uses LLMs to generate
\textbf{whole proofs} at once, without using the intermediate proof state produced by partial proofs~\cite{First23fse, zhang2023getting}. For instance, PALM~\cite{Lu24} samples an LLM for a whole proof attempt and then uses repair mechanisms. The success of whole-proof-based approaches relies on the powerful reasoning abilities of proprietary LLMs, such as OpenAI and Anthropic models, as they must construct proofs that follow viable high-level plans. However, these approaches fail to provide the LLM with access to the intermediate state, which contains important context to aid in reasoning. 

Other approaches perform \textbf{tactic-by-tactic} search, building a proof one step at a time and allowing the LLM to condition the next step generation based on access to the intermediate state~\cite{Yang19, First20oopsla, Sanchez20, Yang23, Blaauwbroek24ICML, Jiang2022Thor}. However, navigating the complex search space of possible proofs is challenging since the local state steers the search without consideration for a high-level proof plan.

These approaches represent two extremes on the spectrum of proof synthesis granularity. 
Our work is a middle ground between these two extremes, incorporating the advantages of whole-proof generation
with leveraging intermediate states where this generation fails. 
To achieve this middle ground, we make a fundamental insight: we can leverage the tree structure of whole-proof attempts to localize valid and invalid \emph{subproofs}. We can keep valid subproofs and recurse on points of failure where we regain access to the state. This insight motivated the design of our tool \tool, a novel LLM-based approach to proof synthesis.

\begin{tcolorbox}

\looseness-1
We present \tool, which uses a \textbf{divide and conquer} algorithm to combine an LLM's high-level planning ability with the advantages of local exploration.
\tool starts by tackling the entire goal, automatically identifying working and broken parts of the proof, reducing the granularity of its attack from the whole proof to individual proof steps \emph{by recursing}, if necessary. 

\end{tcolorbox}

\begin{figure*}[t]
\includegraphics[width=\textwidth]{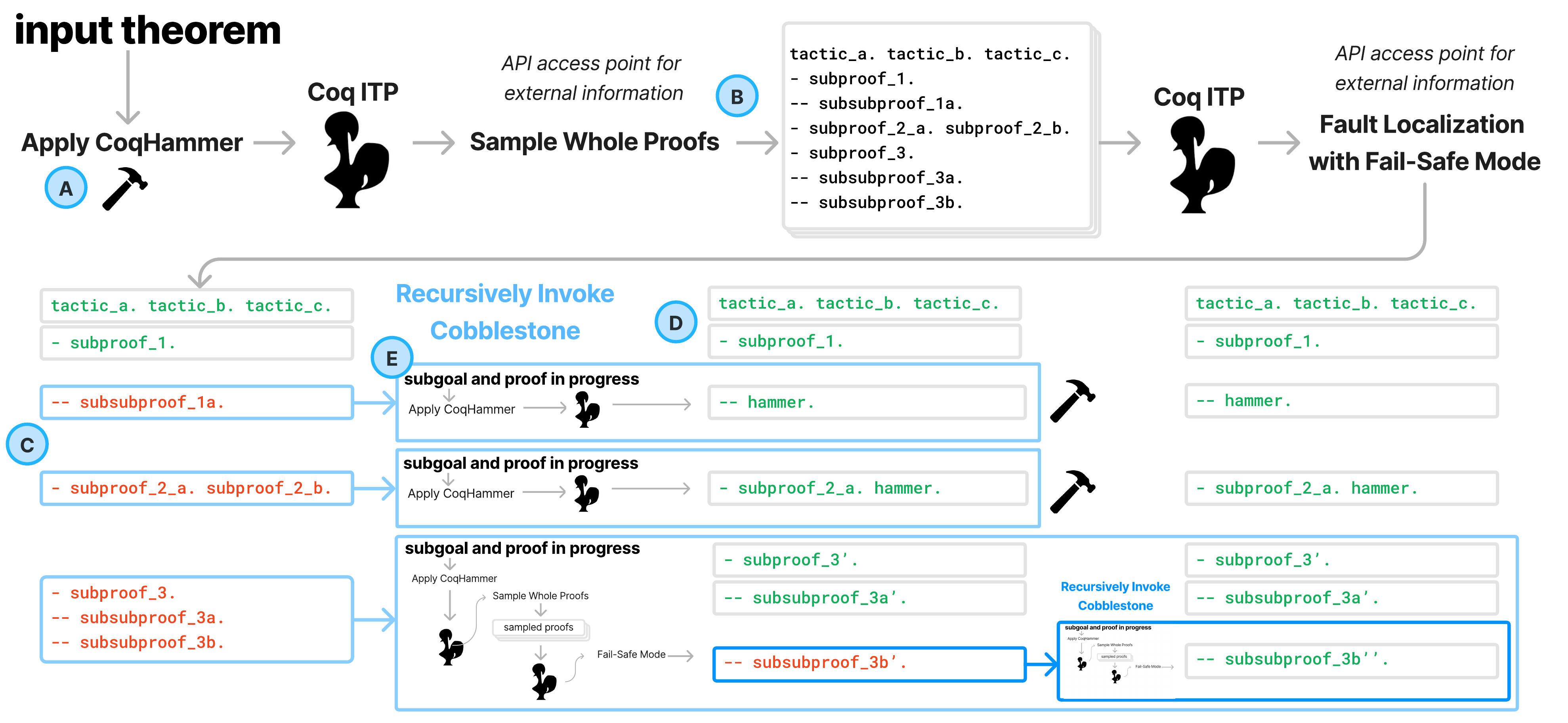}
\caption{
\looseness-1
Given a software property to prove, \tool first attempts to \circledletter[blue]{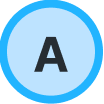}~use \coqhammer to prove that property. If that fails, it \circledletter[blue]{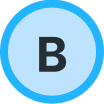}~generates a set of whole proofs, \circledletter[blue]{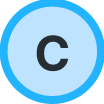}~localizes errors in those proofs, \circledletter[blue]{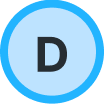}~extracts the working parts, and \circledletter[blue]{E}~recurses on the remaining unproven subgoals to assemble correct proofs. ``API access points'' provide external information as discussed in Section~\ref{sec:rq5}}

\label{fig:high-level-flowchart}
\end{figure*}

\looseness-1
\tool uses an LLM to generate whole proofs, analyzes the proofs using its novel fail-safe execution to decompose the proof into subgoals, and determines which subgoals' subproofs work and which fail. It then attempts to repair the failing subproofs, and, if necessary, recurses on the remaining, unproven subgoals to build a whole, correct proof. 
Because \tool divides the goal into smaller subgoals, it is more effective at using automated theorem proving tools such as hammers; while a hammer may fail to prove the overall goal, it is more likely to succeed on smaller, decomposed goals. Unlike prior work POETRY~\cite{wang2024proving}, \tool does not require fine-tuning an LLM to identify proof substructure and locations to recurse, making it easily extensible to use different LLMs (requiring only minor changes to API calls). 

We implement \tool for the Coq proof assistant, and evaluate it on GPT-4 and two other LLMs. We evaluate on subsets of four benchmarks: 
CoqGym~\cite{Yang19}, a set of open-source Coq projects from GitHub used for evaluating prior proof synthesis tools~\cite{Sanchez20, First20oopsla, First22icse, Sanchez-Stern23toplas, Yang19}; and
\wigderson~\cite{10.1145/3618305.3623600}, \bb~\cite{bb5github}, and \pnv~\cite{pnvgithub}, three recent projects deliberately chosen to be after GPT-4's training cutoff to control for the impact of pretraining data leakage on LLMs.

We compare \tool to prior non-LLM-based proof synthesis tools, LLM-based tools, and LLM baselines of our own creation. 
Prior RNN-based proof synthesis tool, Proverbot9001~\cite{Sanchez20} proves 17\% of the CoqGym subset and 10\% of the \wigderson subset, while
SMT-solver-based CoqHammer~\cite{Czajka18} proves 30\% and 27\%, respectively.
We implement a baseline LLM-based approach that uses chain-of-thought reasoning and show that it proves 25\% and 19\%, respectively.
Meanwhile, \tool, automatically proves 48\% of the CoqGym subset and 38\% of the \wigderson subset. When prior LLM-based tool PALM~\cite{Lu24} is run multiple times to match the token usage of our approach, \tool outperforms it on the \wigderson, \bb, and \pnv subsets, and is highly complementary with it on the CoqGym subset. 
Therefore, \tool complements prior work by proving different theorems they do not, so that together, they prove significantly more.

While completely automated, \tool can also use external input, such as proof structure plans or relevant lemmas from other tools or a proof engineer, further improving proof-synthesis success. 
Combining automated \tool with versions that have access to a proof structure and relevant lemmas can prove 58\% of the CoqGym subset and 55\% of the \wigderson subset.

The main contributions of our work are:
\begin{itemize}[labelwidth=0.7em, labelsep=0.6em, topsep=0ex, itemsep=0ex,
  parsep=0ex, leftmargin=1.5em]

  \item \tool, a novel LLM-based, divide-and-conquer proof-synthesis approach that uses partial successes of failing proofs to produce whole, correct proofs. This serves as a middle-ground strategy between whole-proof and tactic-by-tactic search. 

  \item \emph{Fail-safe mode}, a method for executing a proof to localize proof errors and separate working and failing parts of a proof.
  
  \item A \tool evaluation on four Coq benchmarks and a comparison to state-of-the-art proof synthesis tools showing that \tool consistently proves theorems that prior work cannot, often outright proving more than them.
  
  \item A data-based exploration of how external input from proof engineers or other tools can further improve \tool's success as part of interactive, semi-automated proof-synthesis. 

\end{itemize}

To ensure reproducibility of our results and enable others to build on our work, we make all code, experimental scripts, and data publicly available~\cite{anonCobblestoneReplicationPackage}.

\section{The \tool Approach}
\label{sec:approach}

To formally verify software, in addition to writing the code for a system, an engineer needs to write a mathematical theorem formalizing the desired property, and a high-level proof (proof script) that the property holds. 
 \tool automatically writes the proof script. 
 Given code and a theorem (e.g., \lstinline{in_adj_exists} in Figure~\ref{fig:inadjexists}), \tool synthesizes a proof using a divide and conquer approach, as shown in Figure~\ref{fig:high-level-flowchart}. 
 First, \tool tries synthesizing a proof using \coqhammer (Section~\ref{sec:hammer-repair}). 
 If that fails, \tool samples an LLM for candidate solutions (Section~\ref{sec:prompt}). 
 If none of the candidates prove the theorem entirely, \tool uses a novel fault localization method to isolate errors while noting which subgoals are successfully proven (Section~\ref{sec:error-localization}).
 Next, \tool keeps successful parts of candidate solutions and recurses to find proofs for the yet unproven subgoals (Section~\ref{sec:recursion}), ultimately combining them into a proof of the original theorem.

 While \tool is fully automated, it has extension points (labeled ``API access point for external information'' in Figure~\ref{fig:high-level-flowchart}) that can be used to provide useful additional information. 
 Section~\ref{sec:rq5} will evaluate how this external information can increase \tool's proving power.

\subsection{Illustrative Example}
\label{sec:Coq background}

\begin{figure}
\lstset{style=CoqStyle}
\begin{lstlisting}
Definition adj (g: graph) (i: node) : nodeset :=
  match M.find i g with 
  | Some a => a 
  | None => S.empty end.

Lemma in_adj_exists : forall g i j,
    S.In i (adj g j) -> 
      exists v, M.find j g = Some v /\ S.In i v.
Proof.
  intros g i j H.
  unfold adj in *.
  destruct M.find eqn:E in *.
  - hammer.
  - rewrite SP.FM.empty_iff in H. contradiction.
Qed.
\end{lstlisting}
\caption{A sample theorem
of the \texttt{in\_adj\_exists} property and its human-written proof (lines 9--15).}
\label{fig:inadjexists}
\end{figure}

\noindent
We illustrate our approach using a real-world theorem \texttt{in\_adj\\\_exists}.\footnote{from graph.v in the coq-wigderson project} Figure~\ref{fig:inadjexists} lists a definition (lines~1--4), the theorem statement (lines~6--8), and a human-written proof (lines 9--15).
This theorem states that for all graphs \texttt{g} and all nodes \texttt{i, j}, if \texttt{i} is adjacent to \texttt{j} in \texttt{g}, then \texttt{g}'s structure must contain \texttt{v}.

Coq proofs consist of high-level commands called \emph{tactics}, like ``try induction on \texttt{n}'' or ``try simplifying.'' 
Tactics transform the \emph{proof state}, which contains goals (statements that must be proven) and assumptions that can be used to prove the goals. 
The proof state starts with the initial theorem to be proven.
When tactics transform the proof state to contain no more goals, the \texttt{Qed} command completes the proof.

Figure~\ref{fig:illustrative-example-before-state} shows the proof state before executing the ``\texttt{destruct M.find eqn:E in *}'' tactic on line~12 of Figure~\ref{fig:inadjexists}, which performs case analysis. 
The proof state consists of a single goal and several assumptions, including \texttt{H}, which states that \texttt{i} is a member of \texttt{adj g j}.\footnote{The \texttt{unfold} tactic on line~11 inlines \texttt{adj} in the proof state.} Executing the tactic results in the proof state in Figure~\ref{fig:illustrative-example-after-state}. 
The new proof state \emph{decomposes} the preceding single goal into two \emph{subgoals}. Each subgoal replaces ``\texttt{M.find}'' in the preceding goal with one of its two return values---``\texttt{Some v}'' or ``\texttt{None}''. If both subgoals are proven, then the original theorem is true. 

\begin{figure}
\lstset{style=CoqStyle}
\begin{subfigure}[t]{0.35\linewidth}
\lstset{numbers=none}
\begin{lstlisting}
g: graph
i: S.elt
j: node
H: S.In i (
match M.find j g 
with
| Some a => a
| None => S.empty
end)
---
(1/1)
exists v : nodeset, 
  M.find j g = Some v /\ 
  S.In i v
\end{lstlisting}
\caption{}
\label{fig:illustrative-example-before-state}
\end{subfigure}
\hfill
\begin{subfigure}[t]{0.6\linewidth}
\lstset{numbers=none}
\begin{lstlisting}
g: graph
...
E: M.find j g = Some n
H: S.In i n
---
(1/2)
exists v : nodeset, Some n = Some v 
  /\ S.In i v
===================================
...
H: S.In i S.empty
---
(2/2)
exists v : nodeset,  None = Some v 
  /\ S.In i v
\end{lstlisting}
\caption{}
\label{fig:illustrative-example-after-state}
\end{subfigure}
\caption{
The proof states (a)~before and (b)~after executing \texttt{destruct} on line~12 of Figure~\ref{fig:inadjexists}.
}
\label{fig:illustrative-example}
\end{figure} 

The rest of the proof script is split into two \emph{subproofs}, one for each subgoal, each starting with a bullet (\texttt{-}). 
The first bullet on line~13 invokes \coqhammer using the \texttt{hammer} tactic. \coqhammer\cite{Czajka18} attempts to automatically prove a goal using SMT solvers and proof reconstruction procedures. If \texttt{hammer} succeeds, the goal is proven. Otherwise, the tactic produces an error. 

While \texttt{hammer} is able to find a proof for the first subgoal, it has limitations. For example, \coqhammer cannot perform induction, a key step in many Coq proofs. 
In this case, \texttt{hammer} cannot prove the second subgoal. Instead, after the bullet on line~14 of Figure~\ref{fig:inadjexists}, the proof uses the \texttt{rewrite} tactic, which takes advantage of an already proven lemma (\texttt{SP.FM.empty\_iff}) to change the hypothesis \texttt{H} to enable a proof by contradiction. 
After executing the rest of this proof script, which successfully proves the theorem, the proof state will contain no more goals, indicating that the theorem is proven. 

\subsection{\tool on the Illustrative Example}
\label{sec:example}

\begin{figure}[t]
\includegraphics[width=0.95\linewidth]{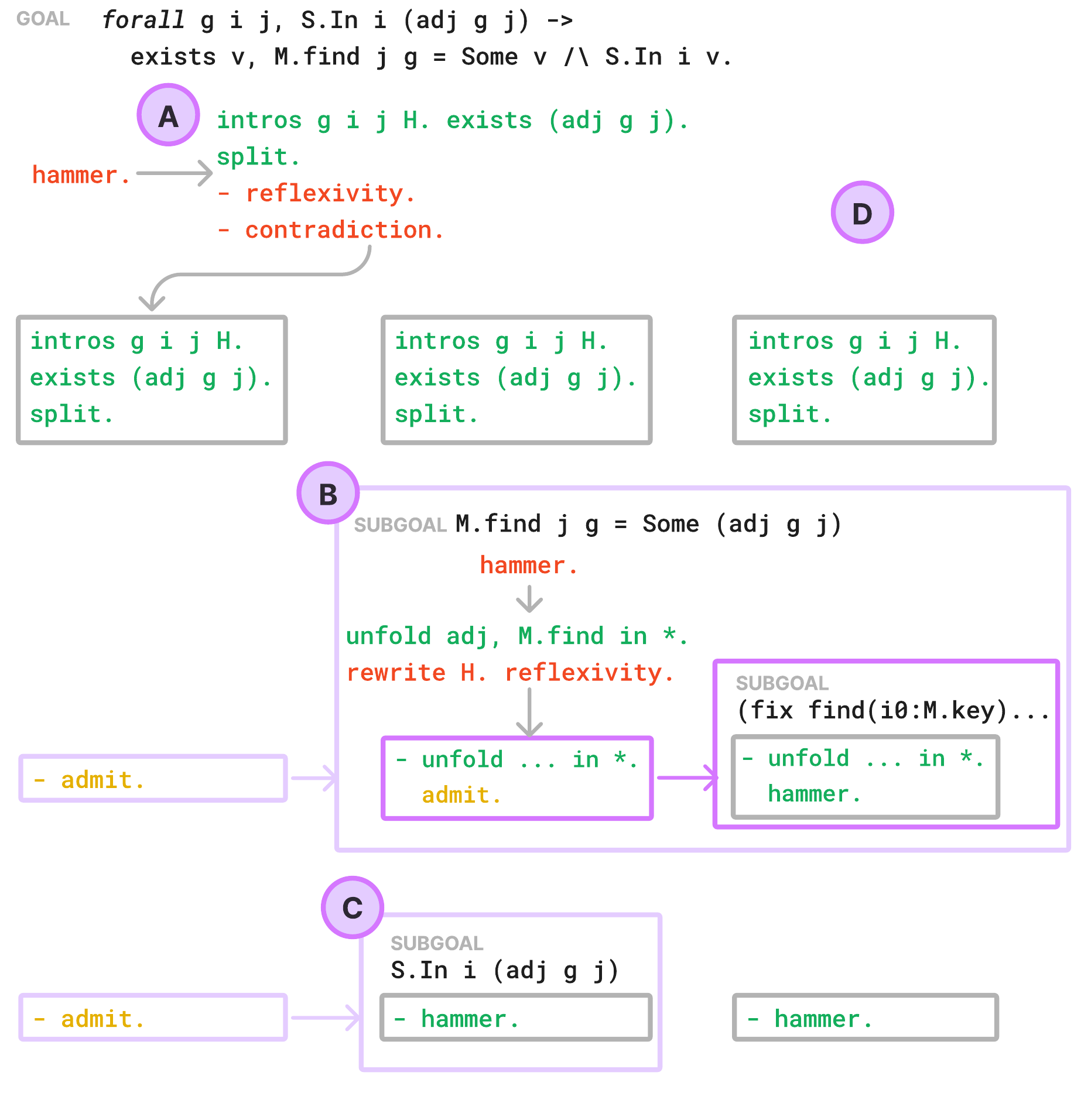}
\vspace{-4mm}
\caption{
Proving \texttt{in\_adj\_exists} with \tool. 
\circledletter[purple]{A} \lstinline{hammer} fails, and an LLM-generated proof contains errors, resulting in recursive calls for each subgoal. 
\circledletter[purple]{B} The first subgoal is proven using a new LLM completion and \lstinline{hammer}. 
\circledletter[purple]{C} The second subgoal is proven using \lstinline{hammer}. 
\circledletter[purple]{D} The final correct proof.
}
\label{fig:concrete-example}
\end{figure}

We now demonstrate how \tool proves \texttt{in\_adj\_exists} automatically.
Figure~\ref{fig:concrete-example} displays the steps \tool takes to divide the proof into smaller pieces and prove each of them, resulting in a correct proof. 
First, \tool attempts to prove \texttt{in\_adj\_exists} in one go, trying \texttt{hammer}, which fails, and then prompting an LLM to generate whole proofs, one of which is shown in \circledletter[purple]{A}. While the LLM's proof starts off with reasonable structure, it contains errors (highlighted in red). 
Normally, Coq will not be able to identify all the errors; Section~\ref{sec:error-localization} describes \tool's novel fault localization mechanism. 
Though the proof in \circledletter[purple]{A} is not correct as-is, it does decompose the initial goal into two subgoals. It may still be a useful part of a correct proof if these subgoals can be proven. However, its correctness can only be determined \emph{retrospectively} after proving these subgoals.

Instead of na{\"i}vely trying to prove \texttt{in\_adj\_exists} again, as prior work has tried~\cite{First23fse}, \tool recurses on each subgoal, attempting to prove it instead (\circledletter[purple]{B} and \circledletter[purple]{C}). In \circledletter[purple]{B}, \tool 
tries using a hammer, which fails, and then prompts an LLM for a proof of the subgoal ``\texttt{M.find j g = Some (adj g j)}''.
This results in a proof with an error, which \tool temporarily fixes by adding the \texttt{admit} tactic. The proof with \texttt{admit}, a \emph{proof in progress}, is passed as an argument to another recursive call. Here, \tool changes the proof in progress to a working proof by replacing \texttt{admit} with the \texttt{hammer} tactic.
In \circledletter[purple]{C}, \tool is able to prove ``\texttt{S.In i (adj g j)}'' using only \texttt{hammer}.

With successful subproofs for each subgoal in hand, \tool combines parts of the proofs in \circledletter[purple]{A}, \circledletter[purple]{B}, and \circledletter[purple]{C} to form a correct proof of the whole theorem, shown in \circledletter[purple]{D}.
Notably, by using LLM completions and its divide-and-conquer approach, \tool can divide and simplify the goals, often reaching subgoals simple enough for \texttt{hammer} to prove. This allows the final proof to integrate portions of two separate LLM completions (for different subgoals), and two invocations of \texttt{hammer}, into a single proof.

\subsection{Applying \coqhammer}
\label{sec:hammer-repair}

\tool combines neural-based proof synthesis and SMT-based proof synthesis by invoking CoqHammer~\cite{Czajka18}.
In Figure~\ref{fig:high-level-flowchart}, the hammer icon (\raisebox{-2pt}{\includegraphics[width=10pt]{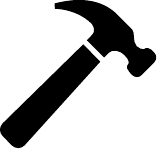}}) shows how \tool uses \coqhammer in two ways. 
First, at the start of \tool's execution,
\tool invokes CoqHammer on the goal, which sometimes produces a working proof. 
Second, in recursive calls, \tool may be passed a \emph{proof in progress}, which consists of working tactics followed by an ``\texttt{admit}''. For example, ``\texttt{unfold adj, M.find in *. admit.}'' in Figure~\ref{fig:concrete-example}~\circledletter[purple]{B} is a proof in progress. \tool replaces \texttt{admit} with \texttt{hammer}, and checks if this proves the subgoal. Attempting to repair proofs in progress like this allows \tool to use CoqHammer on a variety of goals, increasing the frequency of success.

\subsection{Sampling LLM for Whole Proofs}
\label{sec:prompt}

Central to \tool's proof synthesis is using an instruction-tuned LLM~\cite{ouyang2022training} to sample whole proofs.
\tool's prompt consists of two parts---a system message with high-level directions about the task the LLM should perform, and a user message\footnote{``user message'' is terminology used by the OpenAI API. \tool generates these prompts automatically, without a user.} with details specific to the current theorem.

The system message directs the LLM to produce a whole proof for the provided theorem. 
The user message contains 5 types of information: (1)~the theorem statement to prove, (2)~the current proof state, (3)~the definitions for all identifiers mentioned in the theorem statement and proof state that are not in the standard library, (4)~contextual information, and (5)~optional reasoning, which is used as part of chain-of-thought prompting. The user message is formatted as below, with each section preceded by a section header. \tool constructs the user message fully automatically using the coq-serapy library~\cite{Sanchez20} to interact with Coq.

\lstset{numbers=none}
\begin{lstlisting}
[CURRENT THEOREM]
{{theorem statement from Coq ITP}}

[PROOF CONTEXT]
{{current proof context from Coq ITP}}

[DEFINITIONS]
{{definitions from Coq ITP}}

[OTHER PROVEN THEOREMS]
{{proven theorems from Coq ITP and optional oracle}}

[REASONING]
{{optional chain-of-thought reasoning}}
\end{lstlisting}

The contextual information in the prompt can either be empty or contain the theorem statements that were successfully proven before the given theorem in the file.
If the information in the prompt is longer than the LLM's token limit, the contextual information is left-truncated, removing information from the file that is further from the theorem statement. \tool samples the LLM with this prompt, which results in candidate whole proof scripts.

Chain-of-thought~\cite{Wei2022-qv}, a popular prompting technique, prompts an LLM to provide natural language reasoning before generating its final response, and has been shown to help improve LLM performance.
\tool implements chain-of-thought by first prompting the LLM with a modified system message that instructs it to generate reasoning, and then prompting it a second time, with the generated reasoning added to the \lstinline{[REASONING]} section of the user message. When \tool samples an LLM without chain-of-thought prompting, this extra section is omitted.

Prior work 
has shown that the use of diverse data in models can increase the proving power of a proof-synthesis approach~\cite{First22icse}.
Accordingly, to increase variation in \tool's generated proofs, \tool samples the LLM with 4~variations of the same prompt in 2~different dimensions---with and without contextual information and with and without chain-of-thought reasoning. \tool benefits from this approach because the Coq theorem prover can serve as an oracle of proof correctness to pick a correct proof from among multiple samples. 
Our replication package~\cite{anonCobblestoneReplicationPackage} includes all prompts that were used as part of our evaluation. 

\subsection{Localizing Errors with Fail-Safe Mode}
\label{sec:error-localization}

\begin{figure}
\includegraphics[width=.8\columnwidth]{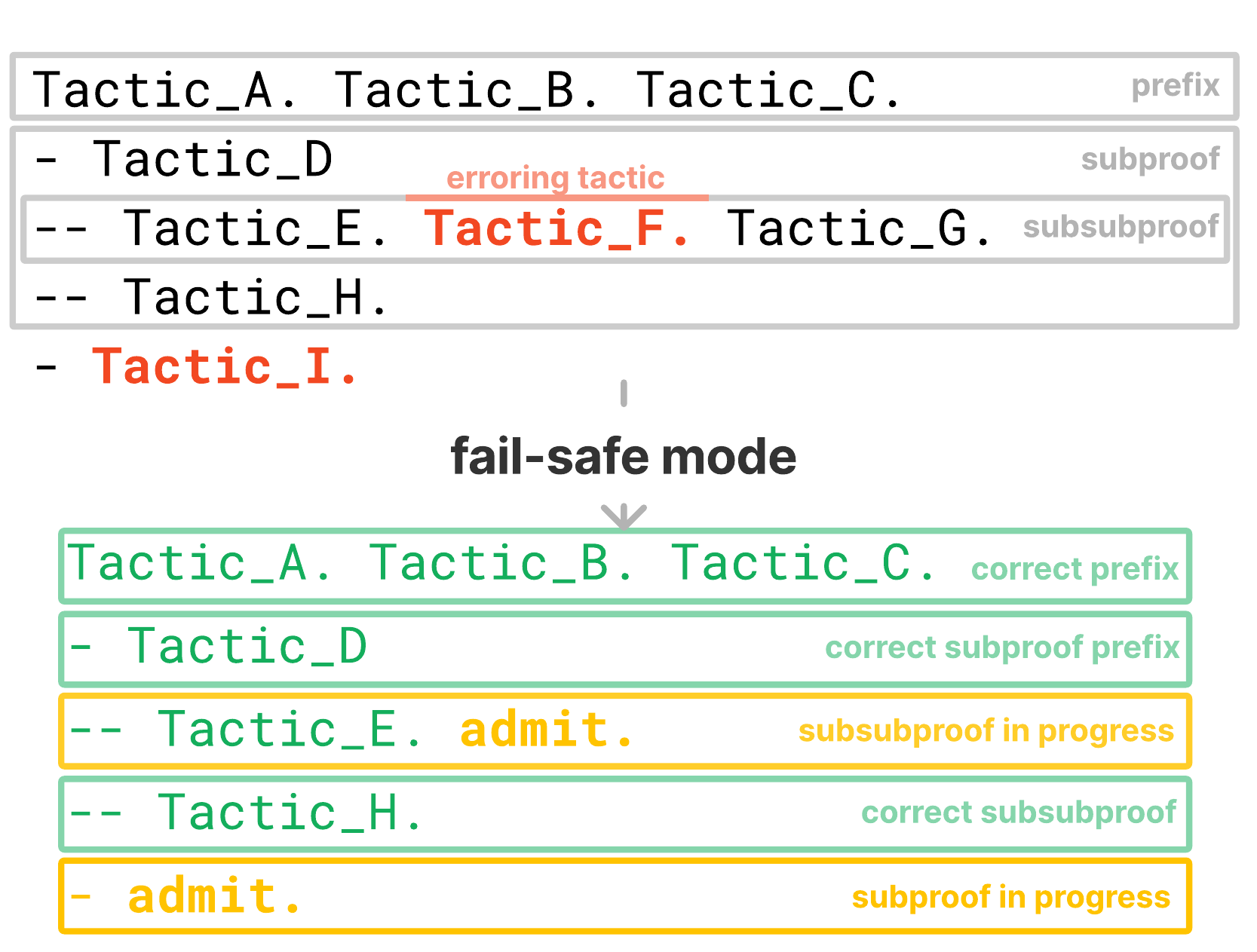}
\caption{An execution of fail-safe mode on an input proof}
\label{fig:failsafe-input-output}
\end{figure}

Once \tool samples proofs from the LLM, it executes each of them in Coq. 
If a proof proves the goal, \tool returns it as-is. 
If none of the sampled proofs are successful, \tool localizes the errors in each proof, identifying which parts are correct, and which parts are not. 
Executing a generated proof na{\"i}vely using the theorem prover stops on the first error, which then misses correct parts that occur after that error.
Instead, we create \emph{fail-safe mode}, a novel way of executing a proof in the theorem prover. 
This approach localizes errors by recursing over the proof's structure, making small edits to allow the execution of proof sections even after an error, and annotating sections as ``correct'' and ``in progress.''
These annotations are implemented via a parallel data structure that keeps track of which subproofs are correct and which are in progress.

Fail-safe mode relies on the fact that many proofs (43\% of proofs from the ChainOfThought evaluation from Section~\ref{sec:rq1}) contain \emph{subproofs}, sections of the proof organized using bullets. 
A proof consists of a \emph{prefix} followed by 0 or more \emph{subproofs}, each of which consists of a bullet followed by a nested proof. 
The last tactic in the prefix decomposes a single goal into multiple subgoals, one for each subproof (recall Section~\ref{sec:Coq background}). Each subproof's bullet focuses Coq on a single subgoal in the proof state; the proof that follows proves that subgoal. Figure~\ref{fig:proof-grammar} formalizes this structure.

\begin{figure}[t]
\small
\begin{align*}
\nonterm{Proof}  &\rightarrow && \nonterm{Prefix}\ \nonterm{Subproof}*\\
\nonterm{Prefix} &\rightarrow && \nonterm{Tactic}+\\
\nonterm{Subproof} &\rightarrow && \nonterm{Bullet}\ \nonterm{Proof}\\
\nonterm{Tactic} &\rightarrow && \T{split.} \mid \T{hammer.} \mid \dots\\
\nonterm{Bullet} &\rightarrow && \T{-} \mid \T{--} \mid \T{---} \mid \dots
\end{align*}
\vspace{-6mm}
\caption{The formal structure of a proof with bullets}
\label{fig:proof-grammar}
\end{figure}

\looseness-1
Note that the definition of subproof is recursive. Each subproof can contain subproofs of its own (subsubproofs). In all figures, we indicate the level of nesting of the subproof by repeating the ``\texttt{-}'' character in bullets (i.e. ``\texttt{-}'' for a subproof, ``\texttt{-{}-}'' for a subsubproof, etc.). 
A proof with no bullets can be expressed solely using a prefix and no subproofs. 
This means that all possible proofs can be expressed using this structure. 
(Section~\ref{sec:rq6} will discuss fail-safe mode's limitations in localizing errors in proofs consisting of only a prefix.)

\begin{algorithm}[H]
\footnotesize
\caption{Fail-safe mode}
\label{algo:failsafe}
\begin{algorithmic}[1]
\Require{A prefix $p$, and 0 or more subproofs $sps$}
\Ensure{the modified proof runs without errors, and the most deeply nested subproofs with errors are annotated as ``in progress''}
\Procedure{RunFailSafe}{$p$, $sps$}
\ForAll{tactics $t$ in $p$} \Comment{First, run $p$}
  \State Run($t$)
  \If{$t$ results in an error}
    \State replace $t$ and everything after it with \texttt{admit}.
    \State annotate $p$ as ``in progress'' \\
    \hskip\algorithmicindent \hskip\algorithmicindent \hskip\algorithmicindent \Return \\
  \EndIf
\EndFor
annotate $p$ as ``correct''

\ForAll{subproofs $sp$ in $sps$} \Comment{then, run $sps$}
  \State{RunFailSafe($sp$.$\textit{prefix}$, $sp$.$\textit{subproofs}$)}
\EndFor
\EndProcedure
\end{algorithmic}
\end{algorithm}

\looseness-1
Algorithm~\ref{algo:failsafe} describes fail-safe mode, which 
takes a proof script broken into prefixes and subproofs as input, modifies it to allow execution after errors, and annotates subproofs, subsubproofs, etc. as ``correct'' or ``in progress.''
First, it executes each tactic in the prefix. 
If a tactic results in an error, that tactic and everything after it is replaced with \texttt{admit}\footnote{Note that inserting \texttt{admit}s into proofs in progress is temporary; \tool will remove these in a recursive call.}, and the prefix is annotated as ``in progress.'' 

If the prefix executes without an error, \tool marks it as correct and fail-safe mode recurses on each of the subproofs, annotating them in the same way. 
Figure~\ref{fig:failsafe-input-output} shows the result of running fail-safe mode. Tactics that run successfully are colored green, tactics that produce errors are colored red, and \texttt{admit}s are colored yellow as they are temporary. The boxes around prefixes and subproofs in the output are annotations, colored yellow if a section is in progress and green if it is correct.
Note that \tool marks the subsubproof starting with \texttt{Tactic\_E} as in progress, but not the subproof starting with \texttt{Tactic\_D} that contains it. 
Because annotations occur as part of a depth-first search, fail-safe mode localizes errors to the most deeply nested subproof possible. 

In addition to the description in Algorithm~\ref{algo:failsafe}, fail-safe mode also handles two complications that arise when executing LLM-generated proofs. 
To understand these complications, consider the example proof that an LLM could generate below and to the left.

\begin{minipage}{.45\linewidth}
\lstset{style=CoqStyle}
\begin{lstlisting}[numbers=none]
Proof.
  intros.
  destruct a.
  apply nonexistent.
  - assumption.
  - auto.
Qed. 
\end{lstlisting}
\end{minipage}
\hfill
\begin{minipage}{.45\linewidth}
\lstset{style=CoqStyle}
\begin{lstlisting}[numbers=none]
Proof.
  intros.
  destruct a.
  - admit.
  - admit.
Qed. 
\end{lstlisting}
\end{minipage}

The first complication is that the proof can encounter an error while executing a prefix tactic \emph{after} it decomposes the original goal. 
Concretely, ``\texttt{destruct a}'' might successfully decompose the goal into subgoals, but then, ``\texttt{apply nonexistent}'' can fail. In this case, instead of just replacing the ``\texttt{apply nonexistent}'' with an \texttt{admit}, fail-safe mode searches for a shorter prefix that leads to multiple subgoals, which we call a \emph{decomposing prefix}. For example, if the sequence ``\texttt{intros. destruct a.}'' leads to multiple subgoals, then fail-safe mode keeps it as the prefix and creates a bullet with an \texttt{admit} placeholder for each subgoal, as shown on the right.

The second complication is that a decomposing tactic can produce more or fewer subgoals than the number of bullets that follow. On the right, there are two bullets that follow ``\texttt{destruct a}''. 
If ``\texttt{destruct a}'' produces fewer subgoals than two, fail-safe mode trims the number of bullets; if ``\texttt{destruct a}'' produces more than two, fail-safe mode adds the appropriate number of bullets with \texttt{admit} placeholders. This guarantees that the number of subproofs that appear syntactically in the proof matches the number of subgoals generated by the prefix.

\subsection{Recursively Invoking \tool}
\label{sec:recursion}

After executing fail-safe mode on each of the four sampled proofs, \tool produces four annotated proofs like the one in Figure~\ref{fig:failsafe-input-output}. 
This example contains two subproofs. 
The first one (starting with \texttt{Tactic\_D}) has two subsubproofs. One is in progress, and the other succeeds. 
The second subproof is also in progress. 

The overall theorem is not considered proven until \tool generates and swaps in correct replacements for \emph{all} the subproofs in progress for one of the four annotated proofs. 
To do this, \tool loops over the annotated proofs. For each proof, \tool makes one recursive call per subgoal in progress. It passes the subgoal and its corresponding subproof in progress (which contains \texttt{admit}) to the recursive call. 
Note that when fail-safe mode annotates a deeply nested subproof as in progress (e.g., the subsubproof that starts with \texttt{Tactic\_E}), it only recurses on the corresponding subgoal, not on any of the subgoals containing it (e.g., the subgoal corresponding to \texttt{Tactic\_D}).
If none of the proofs are able to break the goal into subgoals, \tool calls itself again, retrying with the original theorem as its goal. 
We limit this search procedure using two hyperparameters---\textit{maximum depth}, which limits the level of recursion, and \textit{maximum invocations}, which limits the total number of times \tool can be invoked. These settings are discussed further in Section \ref{sec:experimental-setup}.

Through this greedy depth-first recursion, \tool uses divide-and-conquer, iteratively simplifying the goal and recursing on simpler subgoals, increasing the probability of success. 
This way, \tool breaks down the problem of proving the original theorem into smaller chunks, and leverage progress made in the previous iterations to assist the subsequent ones.

\section{Evaluation}
\label{sec:eval}

We evaluate \tool on 4 datasets of theorems from open-source Coq projects. We compare \tool to 5 prior tools
and two LLM baselines of our own creation.
Our evaluation answers 5 research questions:

\begin{enumerate}[itemindent=0em]
    \item[\textbf{RQ1}:] How does \tool compare to state-of-the-art proof synthesis methods?

    \item[\textbf{RQ2}:] How much does \coqhammer contribute to \tool's performance?

    \item[\textbf{RQ3}:] How much does \tool's search strategy contribute to its performance?
    
    \item[\textbf{RQ4}:] How does external information affect \tool's performance?
    
    \item[\textbf{RQ5}:] How do theorems \tool proves and fails to prove differ?
    
\end{enumerate}

\subsection{Experimental Setup}
\label{sec:experimental-setup}

\mypara{Benchmarks.}
\label{sec:benchmarks}
We construct our evaluation benchmarks from \coqgym's test set of 
theorems~\cite{Yang19}, and from three other open-source projects---\wigderson~\cite{10.1145/3618305.3623600}, \bb~\cite{bb5github}, and \pnv~\cite{pnvgithub}.

\coqgym is a widely used benchmark for evaluating proof-synthesis tools~\cite{Yang19, First20oopsla, First22icse, Sanchez-Stern23toplas}, comprised of 68,501 theorems and their associated human-written proofs across 124 projects.
\coqgym's test set consists of 26 projects with 12,161 theorems, including verdi-raft (distributed software correctness), goedel (G\"odel's 1st incompleteness theorem), and tree-automata (verified algorithms).
Evaluating using this dataset allows for a more fair comparison to prior tools, and assessing tool efficacy across a diverse set of projects.
However, \coqgym was released in 2019, and its projects existed on GitHub before then. 
Because the pre-training cutoff for GPT-4 is September 2021, GPT-4 has potentially seen \coqgym in its pre-training.

To address this concern, we also evaluate using \wigderson~\cite{10.1145/3618305.3623600}, \bb~\cite{bb5github}, and \pnv~\cite{pnvgithub}. The first commit from each of these projects (March 2022, April 2024, and August 2024, respectively) is after the GPT-4 pre-training cutoff.
\wigderson is a formal verification of Wigderson's graph coloring algorithm. It consists of 174 theorems. 
\pnv contains 810 theorems about 
mathematics, including first order logic, Hilbert logic, and boolean algebra.
\bb contains 1446 theorems about busy-beaver values.

Because of the cost of performing inference on LLMs, full-scale evaluations on tens of thousands of theorems are impractical. In fact, the cost of just the LLM usage for our evaluation, including ablation studies, exceeded US\$8K.
Instead, we create four 100-proof subsets of the four benchmark projects, sampling theorems at random.
We call these subsets \coqgymtest, \wigdersontest, \pnvtest, and \bbtest. To keep the cost of ablations low, we evaluate on \pnvtest and \bbtest only for RQ1 (Section~\ref{sec:rq1}).
The total size of this dataset (400 theorems), is in line with other prior work which uses LLMs for proof synthesis~\cite{wanglego, wang2024proving, thakur2024an}, and 300 of the 400 theorems are published after \gptfour's training cutoff.

\mypara{Comparisons to State-of-the-Art and Baselines.}
We compare \tool to 5 prior proof-synthesis tools---%
(1)~\proverbot~\cite{Sanchez20}, an RNN-based neural theorem prover;  
(2)~\coqhammer~\cite{Czajka18}, which uses SMT solvers and proof reconstruction procedures;
(3)~\tactician~\cite{Blaauwbroek20}, which uses an online k-nearest-neighbor algorithm to select tactics during proof search;
(4)~\palm~\cite{Lu24}, which leverages retrieval augmentation (RAG), custom repair prompts, and a backtracking algorithm; and
(5)~\rango~\cite{thompson2025rango}, which fine-tunes an LLM 
and uses RAG 
to include
relevant lemmas and proof scripts. 
\proverbot outperforms other ML-based proof synthesis tools that do not use LLMs, including ASTactic~\cite{Yang19}, TacTok~\cite{First20oopsla}, Diva~\cite{First22icse}, and Passport~\cite{Sanchez-Stern23toplas}.
\proverbot and \coqhammer prove 19.8\% and 26.6\% of theorems, respectively, on CoqGym, 
and prove 17\% and 30\% on \coqgymtest. In a prior ``informal comparison'' 
on a subset of CoqGym, Tactician outperformed \proverbot (except on one project) and sometimes outperformed \coqhammer 
~\cite{Blaauwbroek24ICML}.

For \coqhammer, we use the Z3~\cite{de2008z3}, CVC4~\cite{DBLP:conf/cav/BarrettCDHJKRT11}, Vampire~\cite{10.1007/978-3-642-19835-9_7}, and E~\cite{SCV:CADE-2019} SMT solvers, along with the default timeout settings (a 20 second prover timeout and a 5 second reconstruction timeout), following  prior work~\cite{Sanchez20, Blaauwbroek24ICML}.
We evaluate \tactician's ability to produce a proof fully automatically within a 10-minute timeout. 
Unlike Tactician, \proverbot uses depth limits rather than a time limit, but usually finished its search within 10 minutes as well. 
We run \palm with  \texttt{gpt-4-0613} as its base model. One run of \palm on \coqgymtest used ~3.7K tokens per theorem, 15 times less than the amount for \tool (59K tokens per theorem). To correct for this, we run \palm 15 times for each dataset.
\rango is run with the same setup noted in the paper---using a fine-tuned model with temperature 1.0 and a 10 minute timeout. \rango is run on an NVIDIA RTX 2080 GPU, along with a CPU with 16GB of RAM for proof checking.
We do not compare against COPRA~\cite{thakur2024an}, another LLM-based tool, as it is too cost intensive to run in addition to our other experiments. In the COPRA paper, the authors only ran on a subset of theorems in CompCert that Proverbot9001 is not able to prove, citing budgetary constraints and high cost of evaluation.

We also compare against 2 new baselines, ChainOfThought, and \regen.
ChainOfThought prompts \texttt{gpt-4-0613} 20 times (the same as \tool) in exactly the same way as \tool does (using preceding lemmas and chain-of-thought prompting as described in Section~\ref{sec:prompt}) but only checks the proofs for correctness 
without
further processing. 
\regen also prompts \texttt{gpt-4-0613} up to 20 times in exactly the same way as \tool, but regenerates 
the
proof 
after 
its first failure point. When a proof has a failure, \regen keeps the prefix of the proof that succeeds, and 
continues from the last successful proof state.

\mypara{Metrics.}
\label{sec:metrics}
In line with prior evaluations~\cite{First20oopsla, First22icse, First23fse, Sanchez-Stern23toplas}, we use two metrics: 
\emph{success rate} and \emph{added value}.
The success rate of a tool is the fraction of all theorems the tool is able to prove. 
The added value of tool X over tool Y is the number of new theorems X proves that Y does not, divided by the number of theorems Y proves. To measure the number of theorems a tool can prove in a fixed number of LLM samples, we define the \emph{proven@k} metric. This metric is the sum of the pass@k~\cite{chen2021evaluating} values for each proof in a dataset, and represents the expected number of theorems proven with k samples.

\mypara{Computing Resources.}
Throughout our evaluation, we use OpenAI's \gptfour model (\texttt{gpt-4-0613}), which has a context length of 8,192 tokens. 
In Section~\ref{sec:rq1}, we 
also
use \gptthree (\texttt{gpt-3.5-\\turbo-0125}) and \claude (\texttt{claude-3-opus-20240229}). 
\gptthree has a context length of 16,385. \claude has a context length of 200,000 tokens, but to manage 
inference cost,
we capped its token limit at 10,000. 
We chose these 3 models because their APIs support function calling, a key part of our implementation.
We 
call
\gptthree and \gptfour through the OpenAI Python library. We use the official Anthropic library to 
call
\claude.
For each call to an LLM, we sample with temperature 1.0. 
All experiments are run on machines with Intel Xeon Gold 6230 CPUs and 125GB of memory. We have access to 10 cores from these CPUs, as our experiments were run 
in a virtualized environment. 

We run \tool with a maximum depth of 5, and invoke it up to 5 times. Because each \tool invocation samples an LLM 4 times, \tool uses up to 20 LLM samples in total. 

\subsection{RQ1: How Does \tool Compare to State-of-the-Art Proof Synthesis Methods?}
\label{sec:rq1}

Figure~\ref{tab:rq1} shows success rates for non-LLM tools (\coqhammer, \proverbot, and \tactician), our newly created baselines (ChainOfThought and \regen), LLM-based tools (\palm and \rango), and  \tool. 
\tactician and \rango cannot run on 
some
datasets (reported with N/A) which use unsupported older Coq versions. 

\tool proves 48\% of theorems on \coqgymtest and 38\% of theorems on \wigdersontest. It outperforms each of the non-LLM baselines as well as the \emph{union} of these baselines, All non-LLM, adding 54.5\% additional value in \coqgymtest, and 25.0\%, 40.0\% and 14.6\% added value in \wigdersontest, \pnvtest, and \bbtest respectively.
Across all datasets, \tool proves 46 theorems All non-LLM cannot, All non-LLM proves 14 theorems that \tool cannot, and both prove an overlap of 127 theorems.

On \coqgymtest, running \tool costs \$1.46 and takes in 15.32 minutes, with successful runs costing an average of \$0.23 and completing in 2.54 minutes. On \wigdersontest, running \tool costs \$1.03 and completes in 14.21 minutes, and successful runs cost on average \$0.15 and complete in 3.66 minutes.

\begin{figure}[t]
\centering
\footnotesize
\setlength{\tabcolsep}{1pt}
\begin{tabular}{@{}l@{}cc|cc|cc|cc}
\toprule
                   & success  & added & success  & added  & success  & added & success  & added \\
                   &  rate    & value &  rate    & value  &  rate    & value &  rate    & value \\
\midrule
  & \multicolumn{2}{c}{\coqgymtest} & \multicolumn{2}{c}{\wigdersontest}     & \multicolumn{2}{c}{\pnvtest} & \multicolumn{2}{c}{\bbtest}\\
\midrule
\coqhammer         & 30\% & \phantom{0}66.7\% 
    & 27\%    & \phantom{0}44.4\% 
    & 23\%    & \phantom{0}91.3\% 
    & 27\%    & \phantom{0}59.3\% \\ 
\proverbot         & 17\% &           194.1\%  
    & 10\%    &           280.0\% 
    & 12\%    &           275.0\% 
    & 23\%    & \phantom{0}87.0\% \\ 
\tactician         & N/A  & N/A                
    & 13\%    &           200.0\% 
    & 23\%    &           108.7\% 
    & 22\%    &           113.6\% \\
All non-LLM & 33\% & \phantom{0}54.5\%  & 
    32\%    & \phantom{0}25.0\% 
    & 35\%    & \phantom{0}40.0\% 
    & 41\%    & \phantom{0}14.6\% \\
\midrule
ChainOfThought     & 25\% & \phantom{0}92.0\%  
    & 19\%    &     100.0\%       
    & 25\%    & \phantom{0}84.0\% 
    & 31\%    & \phantom{0}48.4\%\\  
\regen & 31\% &  \phantom{0}61.3\%
& 21\% & 104.8\%
& 33\% & \phantom{0}48.5\%
& 22\% & 104.5\%
\\
\midrule
\palm 1x
& 34\% & \phantom{0}52.9\%
& 16\% & 143.8\% 
& 16\% & 175.0\%
& 23\% & 104.3\%\\
\palm            & 52\% & \phantom{0}17.3\% 
    & 30\% & \phantom{0}40.0\% 
    & 35\% & \phantom{0}42.9\% 
    & 41\% & \phantom{0}26.8\% \\
\palm + 
& 61\% &
& 42\% &
& 50\% &
& 52\% &\\
~~~~\tool \\
\midrule
\rango          & N/A  & N/A 
    & N/A & N/A 
    & 30\% & \phantom{0}56.7\% 
    & 43\% & \phantom{0}25.6\% \\
\rango + 
& N/A &
& N/A &
& 47\% &
& 54\% &\\
~~~~\tool \\
\midrule
\tool              & 48\% &                    & 38\%    &                   & 44\%    &   & 43\%    &   \\  
\midrule
All Together       & 63\% &                    
& 48\%    &                   
& 59\%    &  
& 61\%    &  \\  
\bottomrule
\end{tabular}

\caption{
\tool's performance vs.\ baselines and other proof-synthesis tools. The ``added value'' columns report the percent of theorems \tool proves over each of the other tools. 
}
\label{tab:rq1}
\end{figure}

\tool proves more theorems than 
ChainOfThought and \regen in all datasets, adding value of up to 100.0\% over ChainOfThought and up to 104.8\% over \regen. \tool is also more token-efficient than \regen. On \coqgymtest and \wigdersontest, \tool uses 
far
fewer tokens (59.3K vs.\ 99.3K on \coqgymtest and 42.0K vs.\ 72.3K on \wigdersontest). 
On \pnvtest, \tool uses 51.8K, vs. \regen's 83.7K, and on \bbtest, \tool uses 47.8K vs. \regen's 80.6K.

\begin{figure}

\centering
\footnotesize
\begin{tabular}{lcc|cc|cc}
\toprule
                          & \multicolumn{6}{c}{\coqgymtest} \\
\midrule
                          & \gptfour  & added     & GPT-3.5 & added     & Claude & added     \\
                          &           & value &     & value &      & value \\
\midrule
\coqhammer                & 30\%      & 66.7\% & 30\%      & \phantom{0}40.0\% & 30\%   & 66.0\% \\ 
ChainOfThought            & 25\%      & 92.0\% & 13\%      &           200.0\% & 28\%   & 78.6\% \\ 
ChainOfThought            & 37\%      & 35.1\% & 35\%      & \phantom{0}25.7\% & 41\%   & 24.4\% \\
~~~+\coqhammer  \\
\midrule
\tool                     & 48\%      &       & 37\%      &       & 49\%    &       \\ 
\midrule
\midrule
                          & \multicolumn{6}{c}{\wigdersontest}\\
\midrule
                          & \gptfour  & added     & GPT-3.5 & added & Claude & added     \\
                          &           & value &     & value &      & value \\
\midrule
\coqhammer                & 27\%      & \phantom{0}44.4\% &           27\% & \phantom{0}29.6\%      & 27\%    & 25.9\%      \\ 
ChainOfThought            & 19\%      & 100.0\%           & \phantom{0}8\% & 312.5\%                & 22\%    & 50.0\%      \\ 
ChainOfThought & 31\%      & \phantom{0}25.0\% &           28\% & \phantom{0}25.0\%      & 31\%    & 12.9\%      \\
~~~+\coqhammer  \\
\midrule
\tool                     & 38\%      &       & 33\%      &       & 32\%    &       \\ 
\bottomrule
\end{tabular}
\caption{Performance of \tool and ChainOfThought implemented with different LLMs. ``added value'' is \tool's value added relative to each row.}
\label{tab:otherllms}
\end{figure}

\begin{figure}
\centering
\footnotesize
\begin{tabular}{lcccc}
\toprule
                   & \multicolumn{2}{c}{\coqgymtest} & \multicolumn{2}{c}{\wigdersontest}\\
\midrule
                   & success  & added & success  & added \\
                   &  rate    & value &  rate    & value \\
\midrule
\coqhammer                & 30\% &                & 27\%  &  \\ 
\midrule
ChainOfThought            & 25\% &                & 19\%  & \\ 
ChainOfThought & 37\% & 48.0\%         & 31\%  & \phantom{0}63.2\% \\
~~~~$\cup$ \coqhammer  \\
\midrule
\tool-NoHammer                   & 25\% &                & 16\%  & \\
\tool-NoHammer  & 38\% & 52.0\%         & 30\%  &  87.5\% \\
~~~~$\cup$ \coqhammer  \\
\tool                            & 48\% & 96.0\%         & 38\%  &  137.5\% \\ 
\bottomrule
\end{tabular}

\captionsetup{margin={5cm, 0cm}, width=\linewidth}
\caption{
\coqhammer's use contributes significantly to \tool and ChainOfThought. 
``added value'' is the value of adding \coqhammer.
}
\label{tab:rq2}
\end{figure}

\tool also consistently proves theorems that other LLM-based tools cannot in all datasets, outrightly proving more theorems than these tools in many cases.
\tool proves more theorems than a single run of \palm (\palm 1x in Figure~\ref{tab:rq1}). It also 
outperforms
our \palm benchmark (run 15x) on \wigdersontest, \pnvtest, and \bbtest, proving 8\%, 9\% and 2\% more theorems (the difference in success rates) respectively. Furthermore, in all datasets, \tool proves many theorems that \palm cannot. 
\tool alone prove 47 theorems, \palm alone proves 32 theorems, and both
can
prove an overlap of 126 theorems.
On \coqgymtest and \wigdersontest, \tool uses a similar number of tokens to \palm, with an average of 59.3K vs.\ \palm's 58.1K on \coqgymtest and 42.0K vs.\ \palm's 41.8K on \wigdersontest.
On \pnvtest, \tool uses fewer tokens on average (51.8K vs.\ \palm's 62.0K), while using more 
on \bbtest (47.8K vs.\ \palm's 40.7K).
\tool and \palm's complementarity indicates that \tool may benefit from techniques such as \palm's retrieval 
augmentation (RAG) and LLM-based proof repair.

\tool proves the same number of theorems as \rango on \bbtest, and significantly more on \pnvtest. In both cases, \tool and \rango also demonstrate significant complementarity. On \bbtest, their union proves 54\%, and on \pnvtest, it proves 47\%, more than either tool alone. 
Across all datasets, 
\tool alone proves 28 theorems, \rango alone proves 15, 
and they both prove an overlap of 55 theorems. This indicates that \tool may also benefit from leveraging \rango's RAG. \tool and \rango's token consumption is not straightforward to compare, as \rango uses a 1.3 billion parameter LLM, which is not directly comparable 
to \texttt{gpt-4-0613}, a much larger model.

Since \tool uses an LLM as a black box, it works with any LLM that supports tool use. We modified \tool and ChainOfThought to use \gptthree and \claude, as shown in Figure~\ref{tab:otherllms}.
For all 3 models, \tool outperforms ChainOfThought $\cup$ \coqhammer. However, LLM choice can affect performance relative to \gptfour. \gptthree underperforms, both in ChainOfThought and in \tool. \claude overperforms in ChainOfThought, with mixed results in \tool. 

\begin{tcolorbox}
\textbf{RA1}: \tool outperforms non-LLM based proof generation methods and additional baselines. \tool also performs comparably with LLM-based methods, complementing their results by proving theorems they cannot, and outperforming them in many cases. \tool can be easily adapted to use LLMs other than \gptfour.
\end{tcolorbox}

\subsection{RQ2: How Much Does \coqhammer Contribute to \tool's Performance?}
\label{sec:rq2}

Figure~\ref{tab:rq2} shows that \coqhammer plays an important role. 
On \coqgymtest, \tool-NoHammer (an ablated version of \tool) proves 25\% of theorems, while \tool proves 48\%. 
On \wigdersontest, 16\% vs.\ 38\%. 
On both benchmarks, \coqhammer helps \tool prove 96.0\% and 137.5\% more theorems. 

Even running \coqhammer once improves ChainOfThought's success rate---from 22\% to 36\% for \coqgymtest, and from 17\% to 31\% for \wigdersontest. However, \tool outperforms this simpler combination of the two.
Recall that \tool uses \coqhammer in \emph{each} invocation. 
Another simple combination is to invoke \coqhammer once per theorem, succeeding if it or \tool-NoHammer succeeds.
\tool also outperforms this approach (``\tool-NoHammer $\cup$ \coqhammer''). On \coqgymtest, \tool adds 96.0\% value over \tool-NoHammer, compared to the simpler approach's 52\%. On \wigdersontest, \tool adds 137.5\%, vs.\ the simpler approach's 87.5\%.

\begin{tcolorbox}
\textbf{RA2}: \coqhammer and \tool are significantly complementary. \coqhammer contributes to \tool's performance, but \tool outperforms \coqhammer and other simple combinations of the two. Even without \coqhammer, \tool proves theorems other methods cannot.
\end{tcolorbox}
\vspace{-2mm}

\subsection{$\!\!\!$RQ3: How Much Does \tool's Search Strategy Contribute to Its Performance?}
\label{sec:rq3}

\begin{figure}
\centering
\footnotesize
\begin{tabular}{lcccc}
\toprule
                        & \multicolumn{2}{c}{\wigdersontest} & \multicolumn{2}{c}{\coqgymtest} \\
\cmidrule{2-5}
                        &         success  & added & success & added \\
                        &         rate     & value & rate & value \\                      
\midrule
\coqhammer              &            27\%  &                     & 30\%   &              \\
ChainOfThought          &            19\%  &                     & 25\%   &              \\
\proverbot              &            10\%  &                     & 17\%   &              \\
\tactician              &            13\%  &                     & N/A \\
All Prior Together      &            34\%  &                     & 39\%   &              \\ 
\midrule
TacticByTactic-NoHammer &  \phantom{0}8\%  &  \phantom{0}0.0\%   & 11\%   & \phantom{0}2.6\%           \\
\tool-NoHammer          &            16\%  &  \phantom{0}0.0\%   & 25\%   & \phantom{0}7.7\%           \\
\midrule
TacticByTactic          &            33\%  &            11.8\%   & 40\%   &          20.5\%         \\
\tool                   &            38\%  &            17.6\%   & 48\%   &          30.8\%           \\
\bottomrule
\end{tabular}
\caption{\tool outperforms a simpler search strategy, both with and without \coqhammer. ``added value'' is the value each tool adds over ``All Prior Together''}
\label{tab:rq3}
\end{figure}

\looseness-1
\tool's search strategy is influenced by three key design decisions which we examine in this research question---(1)~its use of whole-proof completions (2)~its use of a recursive divide-and-conquer approach, and (3)~its use of a sampling temperature of 1.0.

\tool's use of whole-proof completions sets it apart from many other proof synthesis systems. By contrast, most neural theorem provers synthesize proofs using tactic-by-tactic search, predicting the next tactic and using a tree search
~\cite{Yang19, First20oopsla, First22icse, Sanchez-Stern23toplas, Sanchez20}. 

\begin{figure}
\includegraphics[width=\linewidth]{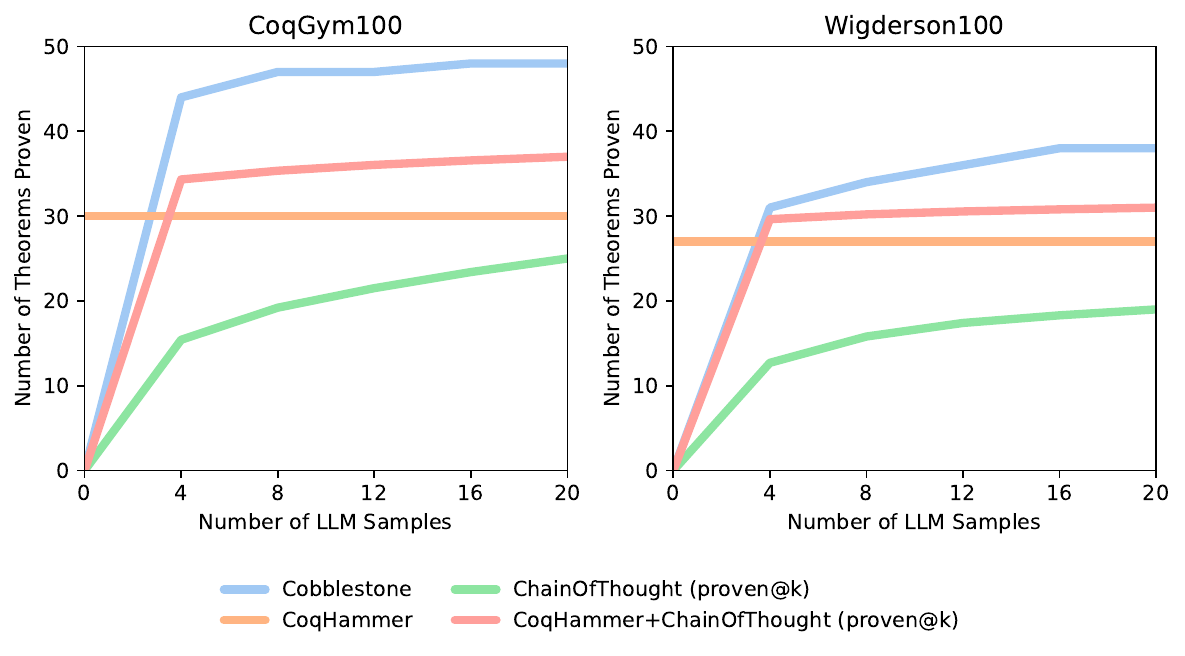}
\vspace{-7mm}
\caption{
Theorems proven vs.\ LLM samples. The y values for ``ChainOfThought'' and ``CoqHammer+ChainOfThought'' use proven@k, detailed in Section~\ref{sec:experimental-setup}.}
\label{chart:num-calls-vs-theorems-proven}
\end{figure}

To measure the effectiveness of whole-proof completions, we implement another tool called TacticByTactic that, like \tool, uses hammer (one for each search step), but only uses \gptfour to predict the next tactic at each search step. We run TacticByTactic, and TacticByTactic-NoHammer, a variant that does not use hammer, allowing up to 20 tactic predictions, a maximum proof depth of 20, and 3 attempts to predict the next tactic at each step. 
As each prediction attempt results in one LLM sample, these tools use the same number of samples as \tool. 

Figure~\ref{tab:rq3} shows that without using \coqhammer, TacticByTactic-NoHammer underperforms all prior tools on \wigdersontest and \coqgymtest. 
With calls to \coqhammer, TacticByTactic performs much better, proving 33\% and 40\% of the theorems in \wigdersontest and \coqgymtest, respectively, and outperforms \tool-NoHammer. Still, \tool outperforms TacticByTactic and adds more value both in \coqgymtest and \wigdersontest.

The second distinguishing feature of \tool's search strategy is its use of a divide-and-conquer approach. As described in Section~\ref{sec:recursion}, when a proof does not work, \tool invokes itself recursively to prove \emph{subgoals} of the original theorem, rather than retrying on the theorem itself. Figure~\ref{chart:num-calls-vs-theorems-proven} shows how this recursive approach contributes to \tool's performance. 

One important datapoint to observe is what happens with a single invocation (which only uses 4 samples). To observe this datapoint, one should look at Figure~\ref{chart:num-calls-vs-theorems-proven}, at the 4 point on the X axis. Even with a single invocation, \tool proves more theorems than the combination of \coqhammer and ChainOfThought can in 20 samples. This is thanks to \tool's error localization (Section~\ref{sec:error-localization}) and careful application of \coqhammer (Section~\ref{sec:hammer-repair}). 

This datapoint is also interesting to observe because it is an ablation that runs fail-safe mode without recursively invoking \tool. This ends up being  similar to PALM's~\cite{Lu24} backtracking algorithm (which also does not make recursive calls).

Invoking \tool recursively further increases the number of theorems it can prove.
On \coqgymtest, \tool is able to prove 44 out of 48 (91.6\%) theorems within 1 invocation, proving an additional 4 (9.4\%) by recursing. On \wigdersontest, it is able to prove 31 out of 38 (81\%) theorems within one invocation, proving 7 (19\%) more by recursing. One possible explanation for this is test leakage in \coqgymtest, which would make more samples of the original theorem ``close enough'' for \tool's to apply \coqhammer and error localization to produce a proof.

\begin{figure}
\vspace{-4mm}
\centering
\footnotesize
\begin{tabular}{lcccc}
\toprule
                   & \multicolumn{2}{c}{\coqgymtest}& \multicolumn{2}{c}{\wigdersontest}\\
\midrule
                   & success  & added & success  & added \\
                   &  rate    & value &  rate    & value \\
\midrule
ChainOfThought-temp0 & 10\%  &  &  9\%  &  \\
ChainOfThought       & 25\%  & 150.0\% & 19\%  &  111.1\% \\ 
\midrule
\tool-temp0          & 48\%  & & 32\%  &  \\
\tool                & 48\%  & \phantom{00}0.0\%  & 38\%  &  \phantom{0}25.0\% \\ 
\end{tabular}
\vspace{-3mm}
\caption{A temperature of 1 may improve performance, and does not harm performance. ``added value'' is value of changing from temperature 0 to 1.
}
\label{tab:temp-0}
\end{figure}

The third distinguishing feature of \tool's search strategy is its use of temperature 1.0 when sampling whole proofs. \tool benefits from sampling diverse proofs (recall Section~\ref{sec:prompt}). A temperature of 1.0 contributes to this diversity. To examine the benefits of a higher temperature, we ran 
\tool and ChainOfThought with the temperature set to 0. As shown in Figure~\ref{tab:temp-0}, in both \wigdersontest and \coqgymtest, ChainOfThought-temp0 proves a subset of the theorems that ChainOfThought is able to prove. In \wigdersontest, \tool proves more theorems than \tool-temp0, with a value added of 25\%, and in \coqgymtest, it proves the same number of theorems.

\begin{tcolorbox}
\looseness-1
\textbf{RA3}: \tool's search strategy outperforms prior tools and an LLM-based tactic-by-tactic search. A single invocation of \tool outperforms a combination of \coqhammer and ChainOfThought, and \tool proves even more theorems when invoked recursively. A temperature of 1.0 can also improve performance.
\end{tcolorbox}

\subsection{RQ4: How Does External Information Affect \tool's Performance?}
\label{sec:rq5}

\looseness-1
\tool can use external information from another tool or a human. 
We created two oracles to provide such information (via the ``API access points'' in Figure~\ref{fig:high-level-flowchart}). 
For each theorem, the \emph{perfect premises} oracle (PerfPrems) knows which proven lemmas the human-written proof uses (\eg, with ``rewrite'' as described in Section~\ref{sec:Coq background}). These are provided in addition to existing context in the \lstinline{[OTHER PROVEN THEOREMS]} section of the prompt in Section~\ref{sec:prompt}.
The \emph{perfect decomposition} oracle (PerfDecomp) 
provides a decomposing prefix (recall Section~\ref{sec:error-localization}) from the human-written proof, breaking the theorem into subgoals for recursive calls. 
Using the PerfPrems oracle is equivalent to asking 
``What lemmas are relevant to proving this theorem?'' and using the PerfDecomp oracle is equivalent to asking ``How would you break down this theorem into smaller goals?''

\looseness-1
We evaluate \tool with access to these oracles, calling the resulting variants \tool-PerfPrems and \tool-PerfDecomp. 
\tool-PerfDecomp has access to both PerfPrems and PerfDecomp oracles.
Figure~\ref{tab:rq5} shows that \tool-PerfPrems outperforms \tool on both datasets, and \tool-PerfDecomp sometimes does even better.
Interestingly, there is some complementarity to the variants, with each proving some theorems none of the others do.
Together, the variants prove 55\% of \coqgymtest and 58\% of \wigdersontest.
The additional information provided to \tool by PerfDecomp does not always result in more theorems proven. 
\tool-PerfPremises and \tool are able to prove 6 theorems in \coqgymtest and 3 theorems in \wigdersontest that \tool-PerfDecomp does not. This indicates that running \tool several times, with different kinds of information in each run, may improve performance.

\begin{figure}
\footnotesize
\vspace{-4mm}
\begin{tabular}{lllll}
\toprule
                     & \multicolumn{2}{c}{\wigdersontest} & \multicolumn{2}{c}{\coqgymtest} \\
\cmidrule(lr){2-3} \cmidrule(lr){4-5}
                     & success  & added & success & added \\
                     & rate     & value & rate    & value \\                      
\midrule
\tool                & 38\% &        & 48\% &  \\
\tool-PerfPrems      & 43\% & 21.1\% & 50\% & 12.5\% \\
\tool-PerfDecomp     & 52\% & 42.1\% & 47\% & 14.6\% \\
All 3 \tool versions & 55\% & 44.7\% & 58\% & 20.8\% \\
\bottomrule
\end{tabular}
\vspace{-2mm}
\caption{External information in terms of relevant lemmas (\tool-PerfPrems) and a breakdown of the proof into subgoals (\tool-PerfDecomp) significantly improves \tool's proving power. ``added value'' is the value each tool adds over \tool. 
}
\label{tab:rq5}
\end{figure}

\begin{tcolorbox}
    \textbf{RA4}: External information, such as useful already-proven lemmas or a decomposition of the theorem into subgoals can significantly increase \tool's proving power, suggesting a promising direction for future research.
\end{tcolorbox}

\subsection{RQ5: How do theorems \tool proves and fails to prove differ?} 
\label{sec:rq6}

To better understand where \tool succeeds, we examine proofs that \tool and its oracle-augmented variants generate. 
To better understand when it fails, we also examine human-written proofs of some theorems it cannot prove. All proofs referenced are available in our replication package\cite{anonCobblestoneReplicationPackage}.

\textbf{Successes:}
\tool's successful proofs range from 1 to 24 tactics long, with the shortest consisting of a single invocation to \coqhammer. Successful proofs with length greater than 1 have an average length of 8.3 tactics, with bullet depth ranging from 1 (no bullets) to 4 (3 levels of bullets). These proofs can consist of 1 to 9 distinct \coqhammer invocations or LLM samples. 
Of the proofs that decomposed their goal, prefixes ranged from 1 to 4 tactics long, with an average of 1.6. Of these prefixes, 41\% decomposed using \lstinline{split}, 36\% used \lstinline{induction}, 18\% used \lstinline{destruct}, and 14\% used \lstinline{apply}.

\looseness-1
The projects in our benchmarks broadly fall into 2 categories---mechanizations of mathematical proofs and formalizations of software. \coqgymtest consists of 12 mathematical projects and 8 software projects with 36 and 64 theorems respectively. \tool proves 61.1\% of \coqgymtest's mathematical theorems and 40.6\% of its software theorems. \wigderson is a software project, and \bb and \pnv are mathematical projects. Including these, our benchmarks contain 236 mathematical and 164 software theorems, with \tool proving 46.2\% and 39.0\% respectively.

Together, \tool-PerfPrems and \tool-\\PerfDecomp generate 26 successful proofs for theorems unproven by \tool (10 in \coqgymtest and 16 in \wigdersontest).
Interestingly, in one of these proofs\footnote{\texttt{votesWithLog\_update\_elections\_data\_timeout} from RefinementSpecLemmas.v in verdi-raft}, the human-written proof relies on 
custom tactics in Coq's Ltac language, while \tool-PerfPrems generates a
proof without custom tactics. In another\footnote{\texttt{Mcardinal\_Scardinal} from graph.v in \wigderson}, \tool asserts a proposition that helps prove the theorem. 

\textbf{Failures:}
\looseness-1
Despite its improvements over the state-of-the-art, \tool and its variants cannot prove 45 theorems from \wigdersontest and 42 from \coqgymtest.
We randomly sampled 14 such theorems (7 from each dataset) to examine manually.

Many of these theorems' human-written proofs are 
much
longer than \tool's proofs. 
\tool's average proof is 10 tactics long, whereas 5 of the 14 unproven theorems have a human-written proof of 20+ tactics.
We also found that some theorems make it difficult for \tool to make partial progress. 
For example, the ground-truth proof of \texttt{request\_VoteReply\_term\_\\sanity\_client\_request}\footnote{from RequestVoteReplyTermSanityProof.v in verdi-raft} 
uses
tactics like \texttt{unfold} and \texttt{apply}, each of which helps prove the theorem, 
without decomposing
the goal.
It is difficult for \tool to generate such proofs as it requires generating a working proof 
without
recursing.
Generalizing our approach to use partial progress that does not involve splitting a goal into subgoals may allow \tool to prove such theorems.

\begin{tcolorbox}
    \textbf{RA5}: \tool's proofs are sometimes shorter than their human-written counterparts, and often successfully leverage its divide and conquer approach. \tool's failures indicate an opportunity to leverage new kinds of partial progress.
\end{tcolorbox}

\subsection{Threats to Validity}
\label{sec:threats}

\looseness-1
All LLM evaluations may suffer from test data leaking into the pretraining dataset, reducing generalizability to unseen data. 
We mitigate this risk by using \wigderson, \bb, and \pnv, whose first GitHub commits are after GPT-4's publicly stated pretraining cutoff date (recall Section~\ref{sec:experimental-setup}).
We evaluate on 400 theorems, in line with other published work~\cite{wanglego, wang2024proving, thakur2024an}, to mitigate cost.

\looseness-1
While our approach is general, our implementation is specific to Coq and uses GPT-4. 
We also evaluate with \gptthree and \claude to show generalizability. When evaluating against other LLM-based approaches such as PALM~\cite{Lu24} and Rango~\cite{thompson2025rango}, we give each tool similar resources \eg by running PALM 15x to control for tokens. However, \tool and Rango's token consumption is not
straightforward to compare, as Rango uses a model with fewer parameters and a different tokenization algorithm.

\looseness-1
Finally, our evaluation only began exploring how external information can aid \tool's proof synthesis using oracles with access to human-written proofs.
The intent of these oracles is to proxy how a human might interact with \tool. But, their performance does not indicate how an interactive, semi-automated approach may perform. 
More research is necessary to study the effects of incomplete and noisy data, as well as user studies to show the potential impact of real, human-provided information. 

\section{Related Work}
\label{sec:related work}

Recent work automating theorem proving in proof assistants has mostly explored three overarching approaches: hammers, machine learning techniques, and combinations of the two. Hammers such as \coqhammer~\cite{Czajka18} and Sledgehammer~\cite{Paulson23} call SMT solvers~\cite{schulz2013system, deMoura08} to construct low-level proofs. They cannot use induction, and thus are limited in what they can prove.
Our evaluation shows that \tool proves many theorems \coqhammer cannot. 

\looseness-1
Machine learning techniques (\ie \emph{neural theorem provers}) use a predictive model learned from existing proofs to predict next steps, and use these predictions to guide a search through the space of proofs~\cite{li2024survey}. They have been built using RNNs~\cite{Huang18, Sanchez20}, LSTMs~\cite{Yang19, First20oopsla, First22icse}, GNNs~\cite{bansal2019learning, Blaauwbroek24ICML}, and LLMs~\cite{Jiang21, Yang23}. LLM-based methods prompt a pretrained model zero-shot or with few-shot examples~\cite{jiang2023draft, zhang2023getting}, fine-tune the model~\cite{First23fse, huang2024mustard}, or use it as an agent~\cite{thakur2023languageagent, wanglego}. 

Hammers and machine learning techniques are complementary~\cite{Jiang2022Thor, First22icse}.
Thor~\cite{Jiang2022Thor} fine-tunes an LLM to learn when to apply Sledgehammer vs. predict a tactic. By contrast, \tool 
samples whole proofs rather than tactics and calls hammer without fine-tuning. Focusing on formalizing math, DraftSketchProve~\cite{jiang2023draft} uses an LLM to translate informal proofs into formal proof sketches with holes, filling the holes with Sledgehammer. Others extend this framework~\cite{zheng2023lyra, wanglego, zhao2023decomposing}. 
LEGO~\cite{wanglego} decomposes theorems into helper lemmas using an LLM, but provides no guarantee that proving them will prove the original theorem; by contrast, \tool uses the theorem prover to decompose goals with such a guarantee. 

LeanDojo~\cite{Yang23} and Magnushammer~\cite{Mikula24} train models to select relevant premises (lemmas/definitions) at each proof step. 
In addition to premises, \rango~\cite{thompson2025rango} also selects similar proof scripts.
\tool uses the preceding lemmas in the file, but could benefit from a premise selection or a proof-script selection model in the same way it benefits from perfect premises provided by an oracle. 

Proof engineers often repair previously working proofs~\cite{Ringer20}.
Symbolic tools can automate some repair~\cite{Ringer21, masci2022proof}. 
Baldur fine-tunes an LLM to repair Isabelle proofs using error messages~\cite{First23fse}. 
Unlike Baldur, which only attempts one repair, \tool is an iterative, divide-and-conquer approach. 
PALM~\cite{Lu24} samples from an LLM and uses repair mechanisms. By contrast, as detailed in Section~\ref{sec:rq3}, our approach is recursive in nature.

\looseness-1
POETRY
fine-tunes an LLM and estimates proof points on which to recurse~\cite{wang2024proving}. By contrast, \tool requires no tuning training data and uses fail-safe mode to precisely identify subgoals that recursion. Adapting \tool to use more powerful models requires only minor changes to its API calls, whereas adapting POETRY is only possible with access to the model's weights, and would require significant resources for fine-tuning.

Formally specifying system properties is complementary to proving them. Prior work has tackled formalizing natural language specifications~\cite{Endres24, Goffi16, Motwani19icse, Zhai20}. 
Once specified, some properties, like fairness~\cite{Galhotra17fse}, can be verified probabilistically~\cite{Thomas19science, Hoag23icse-demo, Metevier19neurips, Giguere22iclr, Albarghouthi17}.

Prompt engineering is key to effective LLM use~\cite{openai2023gpt4, chowdhery2023palm, touvron2023llama}. Frameworks, such as chain-of-thought~\cite{wei2022chain} and others~\cite{yao2023tree, chen2024boosting, besta2023graph},
can elicit reasoning in LLMs~\cite{wei2022chain, wang2022self, zhou2022least, fu2022complexity, chu2023survey}. 
LLMs can be used for quantitative reasoning tasks~\cite{Lewkowycz2022Minerva, azerbayev2023llemma} and programming tasks~\cite{roziere2023code, chen2021evaluating}. 
These advances complement our efforts to generate proofs automatically.

\section{Contributions}
\label{sec:contributions}

\tool is a novel divide-and-conquer method for synthesizing formal verification proofs using LLMs.
\tool significantly outperforms prior
tools and LLM-based baselines, 
proves more complex theorems, and complements other LLM-based tools. 
\tool demonstrates a promising potential for collaborating with humans or other tools to synthesize even more proofs.

\begin{acks}
We thank Zhanna Kaufman and Kyle Thompson for their help executing Proverbot9001 on the \wigderson dataset, and Rango on the \pnv and \bb datasets respectively.
This work is supported by the National Science
Foundation under grants no.\ NSF CCF-1955457, CCF-2210243, and CCF-2220892, and by the Defense Advanced Research Projects Agencies (DARPA) under Contract no.\ HR0011-24-2-0307. 
\end{acks}

\balance
 \bibliographystyle{ACM-Reference-Format}
\bibliography{softeng,laser}

\end{document}